# A Text-Embedding-based Approach to Measure Patent-to-Patent Technological Similarity[*]
– Workflow, Code, and Applications –

Daniel S. Hain[†][φ], Roman Jurowetzki[φ], Tobias Buchmann[ψ], and Patrick Wolf[ψ]

[φ]*AI:Growth Lab, Aalborg University Business School, Denmark*
[ψ]*Centre for Solar Energy and Hydrogen Research Baden-Württemberg (ZSW)*

**Abstract:** This paper describes an efficiently scalable approach to measure technological similarity between patents by combining embedding techniques from natural language processing with nearest-neighbor approximation. Using this methodology we are able to compute existing similarities between all patents, which in turn enables us to represent the whole patent universe as a technological network. We validate both technological signature and similarity in various ways, and demonstrate at the case of electric vehicle technologies their usefulness to measure knowledge flows, map technological change, and create patent quality indicators. Thereby the paper contributes to the growing literature on text-based indicators for patent analysis. We provide thorough documentations of the method, including all code, indicators, and intermediate outputs at `https://github.com/ANONYMEOUS_FOR_REVIEW`).

---

[*]All code necessary to recreate our workflow, indicator creation, and analysis is freely available at `https://github.com/daniel-hain/patent_embedding_research`. We further provide an interactive visualization platform `www.gpxp.org`, which allows exploration and insight creation of all developed indicators and their geographical distribution, similarity networks between countries and technology, and their development over time. All data is also available there for download and own analysis. We hope thereby to spur further research and method development based on semantic indicators of technological development.
Financial support for ZSW's research by BMBF Kopernikus ENavi (FKZ:03SFK4W0)
[†]Corresponding author: dsh@business.aau.dk

# 1 Introduction

Patent data has long been used as a measure of inventive and innovative activity and performance (Griliches, 1990; Pavitt, 1985, 1988; Schmookler, 1966), where last decades have seen a sharp increase in the use of patent-based indicators by academics and policy-makers alike. It has been deployed among others to assess innovation performance of countries (Fu and Yang, 2009; Tong and Davidson, 1994), sectors (Pavitt, 1984), and firms (Ernst, 2001; Hagedoorn and Cloodt, 2003).Its wide availability and coverage across sectors countries and over time makes patent data one of the "go-to" data sources for analysing pattern of technological innovation.

For many applications in patent-based technology analysis, technological similarities between patent pairs or larger patent portfolios represents a key enabling indicator. Patent-to-patent (p2p) technological similarity enables the application of methods and techniques from network analysis to map and understand the structure of technology on various levels of aggregation (e.g. patent, firm, technology, geographical region). A large body of literature leveraged p2p technological similarity indicators to assess knowledge spillovers (Jaffe et al., 1993), and to visualize innovative opportunities (Breschi et al., 2003). It also enables many applications in technology analysis and forecasting, for instance technology mapping and landscaping (e.g. Aharonson and Schilling, 2016; Alstott et al., 2017; Kogler et al., 2013), predicting technology convergence (e.g. San Kim and Sohn, 2020), detecting disruptive technologies (e.g. Zhou et al., 2020), and assessing patent quality (e.g. Arts et al., 2018, 2020).

The wealth of approaches to p2p technological similarity commonly uses one or a combination of the following two data sources: i.) Technology classifications (such as IPC, CPC, e.g., Aharonson and Schilling, 2016; Singh and Marx, 2013)., and ii.) bibliographic data on forward or backward citation of the patents (e.g., Barirani et al., 2013; Huang et al., 2003). Both approaches are subject to a number of limitations. Technology (sub)classes are well suited to categorize technologies, but usually to broad to identify the concrete technological content of a patent (Archibugi and Planta, 1996; Righi and Simcoe, 2019; Thompson and Fox-Kean, 2005), and their static nature limits their usefulness for analysing analyse dynamic phenomena such as the emergence (Kay et al., 2014) or convergence (Preschitschek et al., 2013) of technology. Patent citations are intended to reflect prior art rather than technological content, and citation practices are found to exhibit strategic behavior (Criscuolo and Verspagen, 2008; Lampe, 2012) vary across individuals and jurisdictions (Lemley and Sampat, 2012; Picard and de la Potterie, 2013), and are subject to home- and other forms of bias (Alcacer and Gittelman, 2006; Bacchiocchi and Montobbio, 2010; Griffith et al., 2011; Li, 2014).

Recently, a growing body of research has focused on applying natural language processing (NLP) techniques to utilize the textual components of patent data to derive p2p technological similarity indicators. The textual description of a patents (title, abstract, claims, full text) contains all necessary information to allow domain-educated readers the comprehension of embodied technologies and functionalities. To distill these information in an automated manner, a variety of techniques have been applied, ranging from keyword extraction (Arts et al., 2018, 2020; Gerken and Moehrle, 2012; Kelly et al., 2018; Lee et al., 2009; Moeller and Moehrle, 2015; Yoon, 2008) to the linguistic analysis of subject-action-object (SAO) (Li et al., 2020; Sternitzke and Bergmann, 2009; Yufeng et al., 2016) or ontology (Soo et al., 2006; Taduri et al., 2011) structure. While NLP techniques have already broadened our methodological toolbox for patent analysis, a set of challenges remain. Keyword-based approaches are relatively simple to implement and comprehend, yet their interpretation is complicated by the rich in domain-specific vocabulary, technical and legal jargon, synonyms (the same technology is called different across domains), and antonyms (the same word refers to different technologies across domains) typical for patent text (Beall and Kafadar, 2008; Qi et al., 2020; Tseng et al., 2007). Methods analysing patents' SAO structure or ontology express semantic information better (Yang et al., 2017), but in turn are more time-consuming to calibrate and interpret, and require domain-expert knowledge to do so.

The use of deep learning (LeCun et al., 2015) based embedding techniques (Mikolov et al., 2013; Pennington et al., 2014) has lead to a paradigm shift in NLP, archiving un-preceded performance in many language tasks such as text classification, translation, and semantic search. Such embeddings enable the creation of latent vector representations of textual data which to a large extent preserve the original context and meaning. The potentials of embedding techniques have recently also demonstrated by improving automated patent classification tasks (e.g. Grawe et al., 2017; Kim et al., 2020; Lee and Hsiang, 2020; Li et al., 2018; Risch and Krestel, 2019). This suggests that embedding techniques be used to improve former attempts to create text-based p2p technological similarity measures.

In this paper, we attempt to contribute to the research on patent-based technology mapping by providing a framework leveraging embedding techniques to based on textual data i.) create a patent's technological signature vector, and ii.) to derive measures of patent-to-patent (p2p) technological similarity. We create technological signature vectors for all patents in the PATSTAT database based on their abstract text. We then apply an approximate nearest neighbor search which allows us to process massive data sets and compute p2p similarity measures for the whole universe of patents. We evaluate the validity of the patent's technological signature and derived p2p similarity in multiple way. We first evaluate the quality and usefulness of the derived technological signature for automated technology classification as well as for semantic search. To evaluate the p2p technological similarity measures, we replicate a set stylized facts and compare them to recent non-embedding based approaches. Lastly, at the case of electric vehicle (EV) technologies, we showcase potential research applications in technology mapping, the creation of patent quality indicators, and to identify technological cross-country knowledge flows.

We thereby contribute to the existing efforts to leverage textual data for patent-based technology mapping and forecasting by filling some important gaps. First, we advance research on the creation of p2p similarity measures based on embeddings that is more efficient and less sensitive to domain specific jargon than keyword-based approaches (eg. Arts et al., 2018, 2020; Kelly et al., 2021) Second, by utilizing approximate nearest neighbor search, enabling large-scale cross-technology and-country applications for technology mapping and forecasting without the use of supercomputing infrastructure. Lastly and related, we contribute to the cumulativeness of research in related to p2p similarity, and open and reproducible science more broadly, by sharing well documented code and workflow instructions easing reconstruction and adaptation, as well as all intermediate and final outcomes, such as the p2p similarity measures, word-embeddings and technological signatures (to be found at `https://github.com/ANONYMOUS_FOR_REVIEW`). This eases future research related to the use, creation, advancement, and evaluation of text-based p2p technological similarity similarity indicators.

The remainder of the paper is structured as follows. In section 2, we review the literature on patent-based technology analysis, focusing on approaches to measure p2p technological similarity. We discuss and contrast approaches based on citations, technology classification, and text data. In section 3, we discuss methodological considerations and describe our approach to create a text-based technological signatures of patents, and derive measures of p2p technological similarity. We apply and evaluate these techniques and the obtained results in the following section 4 on all patents to be found in PATSTAT. In section 5 we explore the results of our analysis at the case of EV patents, and demonstrate potential research applications. Finally, section 6 concludes and points towards promising avenues for future research.

# 2 Technological Similarity: Literature Review and State-of-the-Art

## 2.1 Technology class based approaches to measure technological similarity

The existence of distinct technology class labels is a unique feature of patents. Respective class systems, such as the "International Patent Classification" (IPC) or the "Cooperative Patent Classification" (CPC), are taxonomies aiming to capture the entire universe of patented technology (McNamee, 2013), and provide a complex hierarchy of categories to aggregate technological concepts on different levels. Technology classes have been frequently leveraged as a foundation for measuring patent similarity (Zhang et al., 2016). In this regard, especially co-classification analyses have been widely applied (Boyack and Klavans, 2008; Suh, 2017). Such analyses measure the similarity between technology fields by examining the co-occurrence of the classification codes between different patents (Engelsman and van Raan, 1994). Co-classification approaches have the disadvantages that they usually only consider direct overlap and do not take the potential similarity of assigned technology classes into account. More recent approaches therefore derive similarity measures based on the underlying similarity structure of assigned technologies (e.g. Aharonson and Schilling, 2016).

While a large body utilizes technology classifications for patent analysis, they are also subject to a number of critiques. To start with, technology classification systems are said to be too general to satisfy the specific needs of technological forecasting, research planning, technological positioning or strategy making (Archibugi and Planta, 1996). They can be seen as rather vague (Zhang et al., 2016), as researchers are limited to rigid predefined classes. Existing classification shemes do not capture the technological characteristics of an invention sufficiently on class level (Benner and Waldfogel, 2008; Preschitschek et al., 2013; Thompson and Fox-Kean, 2005), where the more fine-grained group- and subgroup-level classes are less stable (WIPO, 2017). On these finer levels, technology classes also tend to display a substantial overlap leading to technologically very similar patents in distant classes (McNamee, 2013). A further challenge in relying on patent classifications schemes is that, as technology changes, similar technology-oriented applications may draw from parents in different hierarchical categories (Kay et al., 2014). Classification systems define new technologies based on already existing technologies or their combinations, leading to uncertainty regarding the actual accuracy of patent-class fit. Previous research also found broad heterogeneity across different patent classification schemes in terms of their weighting of technological functionality versus industry-specific applicability (Adams, 2001), and concludes that in practice several classifications should be applied and considered to provide a more complete picture (Wolter, 2012). Even within a single classification scheme, across countries important features may be lost in the process of classifying technical ideas described in the patent in a common language (Meguro and Osabe, 2019). Depending on human judgement, sometimes they are also just poorly assigned by the respective authorities (Leydesdorff, 2008).

## 2.2 Citation-based approaches used to measure technological similarity

Patent citation represent another popular data source measure p2p similarity, where we can distinguish between three main approaches: i.) co-citation, ii.) bibliographic coupling, iii.) direct or indirect citation.

Co-citation approaches measure p2p similarity using the amount of shared forward citations of two patents Yan and Luo (2017), following the intuition that such co-citations signal overlap in patents' technologies or application. Leveraging forward-citation data, co-citation based measures are only available ex-post patent application, once sufficient citations have accumulated.

In contrast, bibliographic coupling approaches measure p2p similarity by amount of joint backward citations Yan and Luo (2017), assuming that overlap in the patents' references indicates both to be build and utilizing on similar technologies (Von Wartburg et al., 2005). Since a patent's backward citations are available at the time of application, derived measures can be used to create ex-ante indicators.

The third approach relies on using direct/indirect citation (paths) to measure p2p similarity. It calculates a compound similarity matrix based on a patent citation network, represented by a direct similarity matrix, and resulting indirect similarity matrices (Wu et al., 2010). The approach has advantages compared to approaches based on co-citation and bibliographic coupling as it provides more complete information, allowing a more precise assessment of technological distance (Rodriguez et al., 2015; Wu et al., 2010).

Generally, citation based approaches to derive similarity measures are subject to a set of shortcomings. Citation practices differ across patent authorities (Picard and de la Potterie, 2013) and even examiners (Lemley and Sampat, 2012). Further, applicants may withhold citations to prior art for strategic reasons (Criscuolo and Verspagen, 2008; Lampe, 2012), or just not provide useful citations (Cotropia et al., 2013) which instead have to be added by the examiner (Alcácer et al., 2009). It can also not be taken for granted that examiners are willing and able to refer to all relevant prior art. Furthermore, both inventors (Griffith et al., 2011; Li, 2014) as well as patent examiners (Bacchiocchi and Montobbio, 2010) are subject to home bias, making them more likely to cite patents within higher geographical, social, or institutional proximity. Furthermore, despite innovation in the field of patent databases and search technology, prior art discovery and examination remains a challenging activity, and patent search reports are also of varying quality and information richness (Michel and Bettels, 2001) and still require substantial domain expertise to be used correctly. Consequently, the absence of citations is not a sufficient condition for the absence of similarity, and p2p similarity measures based on citation data is likely to result in an substantial amount of false negatives.

## 2.3 Natural language-based approaches to measure technological similarity

Recently, researchers have started leveraging text-based approaches (based on e.g. patent title, abstract, keywords, or claims) attempting to describe the technologies embodied in a patent in a more nuanced way to measure p2p similarity, and map technology landscapes and evolution. In this regard, different methodologies have been developed, which cover i.) keyword-based approaches, ii.) the analysis of the SAO-structure, iii.) ontology-based analysis, and iv.) machine learning and deep learning based approaches.

Keyword-based methods are based on keyword frequency and co-occurrence measures. This approach has often been used in the past due to its simplicity and straightforwardness (Kelly et al., 2018; Lee et al., 2009; Moeller and Moehrle, 2015; Yoon, 2008). To measure p2p similarity, either raw (Arts et al., 2018) or "term frequency-inverse document frequency" (TFIDF, Salton and Buckley (1988)) weighted (Arts et al., 2020) keyword or multi-word (n-gram) (Gerken and Moehrle, 2012). keyword

co-occurrence has been used. However, a major shortcoming of keyword-based approaches is that it fails to reflect relationships among related concepts described by different words. This is particularly true for documents with rich in domain-specific vocabulary, technical and legal jargon, synonyms (the same technology is called different across domains), and homonyms (the same word refers to different technologies across domains), as patents usually are (Beall and Kafadar, 2008; Qi et al., 2020; Tseng et al., 2007).

Using the SAO-methodology, some studies apply the subject−action−object (SAO) structure of patent texts as their semantic representation and aim at introducing more grammatical and meaning structure (Yang et al., 2017). Regarding the calculation of patent similarities, the method has often been combined with additional models and indicators, such as vector "(Visual Syntax Method)" (VSM) models (Yufeng et al., 2016), the Sørensen-Dice index (Li et al., 2020) or Jaccard and Cosine index (Sternitzke and Bergmann, 2009). While this approach is able to take a deeper look at semantics in texts, a major drawback can be seen in the focus on only a small proportion of the available words and therefore a possible miss of relevant information.

Another methodology that recently has gained attention is the analysis of patent texts by their ontology. The approach bases on the construction of an ontology which describes the concepts and respective relations for a specific domain. Based on this domain then a semantic annotation on patent texts is performed. Examples are the analysis system proposed by Taduri et al. (2011) and Soo et al. (2006). While providing strongly semantic modelling, the ontology-based approach is highly labour-intensive and context-sensitive which makes it hard to apply the procedure on a broader scope of patents.

While machine learning (and later deep learning) based approaches for text analysis and classification have been around since the 1990s (Hayes and Weinstein, 1990; Newman, 1998), they have only recently found growing attention in patent analysis, mainly for automated patent technology classification. They are able to map the complex relationships of unstructured texts, and yielded promising results when applied to patent text (Li, 2018). Tran and Kavuluru (2017) explore text data and machine learning-based classification in the context of the Cooperative Patent Classification (CPC) system. Such exercises build on earlier work of automated patent classification for IPC classes (Fall et al., 2003). Lately, the use of deep learning based embedding techniques has lead to a general paradigm shift in NLP and replaced a large set of keyword-based and linguistic approaches to language modelling. Embedding techniques utilize deep neural networks trained on large amounts of text data create high-dimensional vector representations of words or documents, which preserve its original meaning and context. Embedding techniques have found applications in patent analysis, mainly for deep learning based automatic patent classification. For instance, Grawe et al. (2017) compute word embeddings in order to develop a patent text classifier. Li et al. (2018) developed a deep learning-based patent classification algorithm, which bases on convolutional neural networks and word vector embeddings. Chen et al. (2020) developed a method for extracting semantic information from patent texts by using deep learning, and Lee and Hsiang (2020) and Bekamiri et al. (2021) apply currently state-of-the-art language models (BERT), and achieve a currently unpreceeded performance in text-based patent classification.

The ability of embedding-based machine learning models to predict a patent's assigned technology classes with a high accuracy indicates that embedding techniques indeed preserve the technological content and context of a patent. While proven to perform well for the task of classifying patents with respect to their technology class, they do not provide similarity measures between patents, or suggest a workflow of how to do so. In contrast, applications using embedding techniques to derive and evaluate p2p technological similarity measures are scarce, face computational challenges, and are generally not reproducible. Firstly, (Younge and Kuhn, 2016) deploy massive distributed computing power are able to create similarity measures for all patents. Most recently, Whalen et al. (2020) develop patent similarity dataset based on a vector space model that contains similarity scores of US utility patents. Overall, computational bottlenecks have resulted in little progress so far, where the few existing approaches can either provide similarity measures only for subsets of patents, or require massive computational power and are therefore not accessible for most researchers.

In summary, while just recently gaining popularity, text-based approaches to patent analysis already have a long history. In the past mostly keyword-based approaches have been utilized, but more recently the particularities of patent text such as the frequent use of synonyms and domain-specific technical jargon have led to a gradual replacement by embedding-based NLP methods. These have been applied successfully for patent classification tasks, yet less so for measuring technological similarity, and when then on limited subsets of patents and/or by using computational power not accessible for the common researcher. Reasons therefore are among others found in the massive computational demands when creating large similarity matrices. Therefore, we face limitations in utilizing such similarity to map and analyze global technology development.

## 2.4 Summary on approaches to measure technological distance

Table 1 summarises common approaches to measure technological similarity based on patent data. It delineates previous work related to patent similarity measures and how modern approaches based on word vectors and text embeddings can improve the quality of these measures. In particular, while drawing on earlier research on text vectorization and embeddings, we are among the first to apply this approach to derive p2p technological similarity measures, and provide a reproducible workflow for doing so. Additionally, we developed a method that delivers relatively high accuracy combined with time efficiency and scalability which allows application to very large numbers of patent pairs without the requirement of supercomputer power.

Table 1: General overview on methods for patent similarity analysis

| Method | | Description | Characteristics | | Example |
|---|---|---|---|---|---|
| | | | Advantages | Limitations | |
| Class-based measures | Co-classification / Classification overlap | Measures distance between technology fields by examining the co-occurrence of the classification codes (IPC/CPC) between different patents | • Easy access to data from well-structured database<br>• Relatively easy to apply | • Comparison is limited to predefined classes of patents<br>• Classification systems are often too coarse | Engelsman and van Raan (1994); Yan and Luo (2017) |
| Citation-based measures | Direct / indirect | Direct or indirect citation (paths) between patents are used to derive similarity indicators | • Easy access to data from well-structured database<br>• Relatively easy to apply | • Subject to strategic citation and other citation bias | Rodriguez et al. (2015); Wu et al. (2010) |
| | Co-citation | Measures the knowledge distance by calculating the amount of shared forward citations of two patents | • Easy access to data from well-structured database | • Not applicable for most recent patents as enough forward citations are needed<br>• Not applicable if patent hasnäÄŹt been cited | Mowery et al. (1998); Rothaermel and Boeker (2008) |
| | Bibliographic coupling | Measures the knowledge distance by calculating the amount of shared backward citations/references of two patents | • Easy access to data from well-structured database | • Patent references are often not fully comprehensive | Garfield (1966); Von Wartburg et al. (2005) |
| Text-based measures | SAO-based | Looks at the subjectâĂŞverbâĂŞobject structures of patent texts to extract expressions of meaning. The similarity of patents can than be assessed by the similarity of the SAO-structures | • Includes grammatical structure and meaning<br>• Allows to take a deeper look at semantics in texts | • Still considers only a small proportion of the available words (possible miss of relevant information)<br>• Often treats each identified relationship as equally important, which does not necessarily provide an accurate measure of patent similarity | Li et al. (2020); Wang et al. (2019); Yufeng et al. (2016); Sternitzke and Bergmann (2009) |
| | Keyword-based | Caompares patent texts by frequency and/or co-occurrence of specified keywords | • Established approach<br>• Straightforward in application | • Solely relies on single word or few single words<br>• Fails to reflect the relationship between concepts<br>• Prone to false negatives when synonyms or technology specific jargon is used | Kelly et al. (2018); Moeller and Moehrle (2015); Yoon (2008) |
| | Ontology-based | Compares patents by representing the textual content (abstracts, claims, etc.) of a patent as an ontology | • Approach provides a strongly semantic modelling | • Highly labour-intensive<br>• Highly context-sensitive<br>• Barely usable for analyses with a very broad scope | Taduri et al. (2011); Soo et al. (2006); |
| | Embedding-based / vector-based | Transforms text into a numeric vector by using a vector space model. Texts can then be compared by the similarity of the vectors | • Higher accuracy<br>• Able to map complex relationships of unstructured texts<br>• Able to handle large amounts of unstructured Text | • Not straightforward<br>• Often requires either extensive hardware use or a reduction of the number of patents | Whalen et al. (2020); Grawe et al. (2017); Younge and Kuhn (2016); Li et al. (2018) |

# 3 An Embedding Approach to Create P2P Technological Similarity Measures in Patent Data

## 3.1 General logic

Building on recent research in text-based patent analysis as well as methodological advances in the broader NLP field, we apply embedding techniques to patent text. We thereby aim at capturing the patent's technological features and content in a high-dimensional numeric vector, which can be interpreted as the patent's "technological signature". We argue that this technological signature not only represents a more appropriate and nuanced characterization of a patent's technological content, but also a more suitable foundation for p2p similarity measures. In the following, we describe the techniques, parameters, and general logic behind every step of the proposed indicator computation in detail. Figure 1 illustrates the proposed techniques and workflow to create p2p technological similarity indicator based on their textual data, which can be used for a variety of analyses and indicator creation.

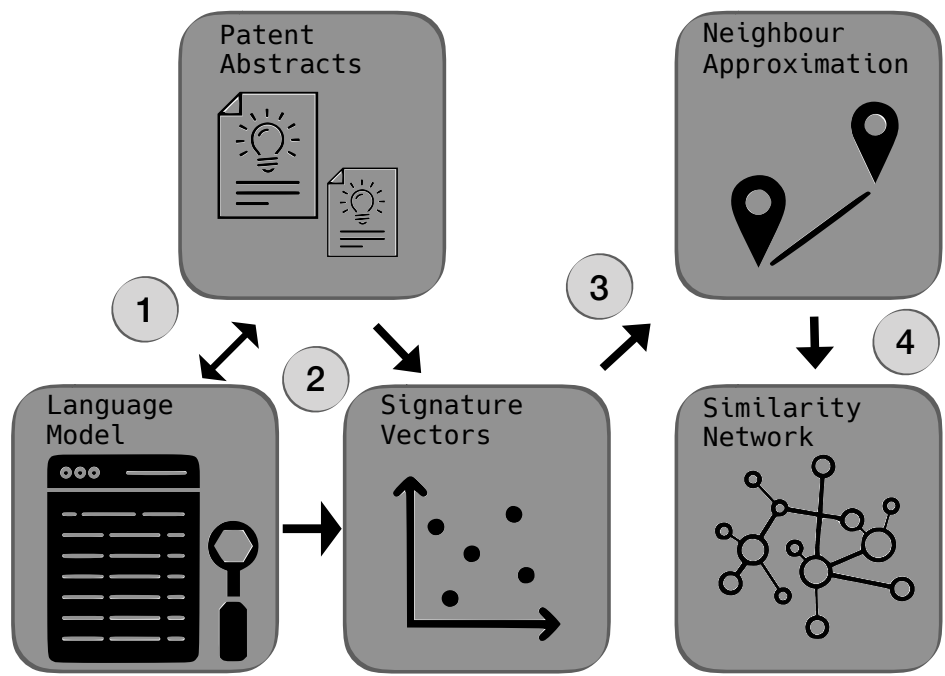

Figure 1: Preprocessing pipeline

## 3.2 From patent text to technological signature vector: Creating document embeddings

To provide such a text-based technological signature, we leverage word and document embedding techniques, which represent a methodological breakthrough that has revolutionized NLP research throughout the last decade. In word embedding models such as Word2Vec (Mikolov et al., 2013) or GloVe (Pennington et al., 2014), word meanings are learned from the context that surrounds the term rather than merely within-document co-occurrence. This principle has been famously summarized as "you shall know a word by the company it keeps" (Firth, 1957). Training of such models on large datasets allows to account for syntax and to extract higher-level meaning structures for terms. Embedding techniques have during the last decade become one of the most promising technique within NLP, and applied for a broad range of application such as semantic search and text classification. Lately, it has been demonstrated that embedding techniques are indeed able to infer and encode complex relationships from textual data, such as the relationships between chemical molecules (cf. Tshitoyan et al., 2019). Embedding techniques help us to overcome limits of keyword-based approaches for patent data such as synonyms, homonyms, disciplinary jargon, and changing meaning over time (Beall and Kafadar, 2008; Qi et al., 2020; Tseng et al., 2007).

Summing and averaging such word vectors has proven to generate good document representations that are able to deal with some of the idiosyncrasies of natural language that simpler models were not able to account for. To calculate document embeddings, we first train a custom word embedding model using the Word2Vec approach on approximately 48 million English patent abstracts found in PATSTAT. We train this custom model instead of using generic pretrained word embeddings due to the arguably specific language found in patent descriptions. In addition, we train a simple TF-IDF model on the whole corpus of patent abstracts. Abstract embeddings are obtained by taking the dot product of the word-embedding matrix with the dense TF-IDF weighted Bag-of-Word representations of the abstracts. As a result, we obtain a 300-dimensional patent signature vector that can be used for further calculations. While TF-IDF vectors, sometimes used in similar work are often sparse high-dimensional objects (1 dimension per term in the vocabulary or combinations of terms) Arts et al. (e.g. 2020) use 1,362,971 dimensions, our 300-dimensional vectors are in comparison relatively compact allowing efficient computation and are easy to share and reuse in different contexts. Below, we describe how we further increase efficiency by applying approximate nearest neighbor search and encapsulating the embedding within a "search object". [1]

## 3.3 From technological signature to p2p similarity: Approximate nearest neighbor search

After creating a technological signature vector for every patent, we attempt to measure p2p technological similarity the the universe of existing patents. For smaller datasets, this can be done with a standard k-nearest neighbor (KNN) search where a similarity score (e.g. euclidean or cosine distance) for each pair of observation is calculated. However, for our population of circa 48 million patents, this would not be possible with reasonable effort, since it would require the calculation of a matrix of size $n * (n - 1)$.[2]

Approximate nearest neighbors computation is an active area of research in machine learning and one of the common approaches to this problem is using k-d trees that partition the space to reduce the required number of distance calculations. Search of nearest neighbors is then performed by traversing the resulting tree structure. Utilizing such an approach can reduce complexity to $O[DNlog(N)]$ and more. In our case, this leads to an efficiency increase by a factor of at least $1.12e^4$.[3] We in the next step calculate the cosine similarity between focal patent and all other patents to be found in neighbouring leaves of the search tree:

$$sim^{cosine}(x, y) = \frac{x^T y}{||x|| \cdot ||y||} \quad (1)$$

---

[1] Python's Gensim library (Rehurek, 2010) is used for the training https://radimrehurek.com/gensim/. Bi-grams occurring over 500 times are aggregated into individual tokens before training. The Word2Vec model runs over 5 iterations, using a window of 8 words, 300 dimensions for the target vectors, terms occurring less than 20 times are not considered. The dimensionality of 300 is a common conservative upper bound parameter.

[2] An example of the data and compute intensity of such an approach is provided by Younge and Kuhn (2016), who produced a patent similarity matrix with 14 trillion entries by using thousands of distributed CPUs for months to do so.

[3] We utilize the efficient *annoy* (Approximate Nearest Neighbor Oh Yeah!, Bernhardsson (2017)). Documentation of the *annoy* package can be found at https://github.com/spotify/annoy implementation that constructs a forest of trees (100) using random projections.

We discard patents-pairs with a cosine similarity below the threshold of 0.65 in the realational evaluation (4.3) in order to create an appropriate level of sparsity to avoid the problem of storing and processing extremely large matrices and reduce the analysis to a space where similarity can be meaningfully measured.[4]

Overall, we argue that this approach, although combining several techniques, has its strength in being extremely scalable and efficient. In comparison to many of the other techniques proposed in the literature our patent vector representations can be created on readily accessible hardware. The resulting approximate nearest neighbor search object is a self-contained file that includes all embeddings and can be used from disk on any modern notebook computer. Here, n similar patents are identified among the full sample of over 48 million patents within on average 60ms (wall time). For large computational tasks like in our case where the aim is to construct a similarity network for all patents the object can be preloaded into memory bringing calculation time down to under 0.5ms.

### 3.4 Data

#### 3.4.1 Textual data

For text-based patent analysis, potentially usable text information likely to contain technological information are a patent's i.) abstract, ii.) claims, iii.) full text description, or a combination thereof. These text segments are drafted for a different purpose and subject to different requirements, therefore will display different properties when used for a text-based analysis. Previous research indicates that simply combining different textual sources tends to decrease rather than improve the information gain as compared to a single text source analysis ([Cetintas and Si, 2012](#)).

---

[4] The chosen threshold of 0.65 is based on the comparison of patent pairs at certain similarity threshold, where we informed by domain experts decided where patents still contain enough meaningful relatedness allowing an interpretation.

Intuitively, a patent's full text description could be assumed to contain the largest and most nuanced set of information regarding its embodied functionality and technology. However, patent full text descriptions are compared to abstracts or claims less subject to scrutiny and regulations regarding their format, therefore diverge considerably in terms of length, style, and clarity. Due to the resulting text heterogeneity and increased signal-to-noise ratio, previous research has predominantly favored the use of abstract or claims text ([Noh et al., 2015](#)). Among those, a considerable body of research utilizes patent claims text for patent classification (e.g. [Lee and Hsiang, 2020](#)), and to infer patent quality indicators such as novelty (e.g. [Marco et al., 2019](#)). As data source for the creation of p2p technological similarity measures, an abstract arguably contains a broader overview on the embodied technologies. An abstract usually not only contains all of the inventions' features given in the patent claims, but further information on the technical field and examples of possible uses. Further, claims tend to highlight legally protectable difference rather than similarity. Abstracts are therefore better suited when aiming at technological similarity in general, while claims are for example well suited when specifically looking at patent infringement. Abstracts are said to "communicate the technical description in a concise and straightforward manner, avoiding unnecessary words that may increase noise in the extraction process" ([Tshitoyan et al., 2019](#), p. 98). Lastly, abstracts in comparison to claims available widely available across abstracts and jurisdictions, easing attempts to carry out global technology mapping.

### 3.4.2 Patent data

The patent data we used for our study was retrieved from the EPO's Worldwide Patent Statistical Database (PATSTAT, Autumn 2018 edition) which covers bibliographic patent data from more than 100 patent offices of developed as well as developing countries over a period of several decades.

We in a first step create technological-signature embeddings for all patents where English-language abstracts are available (ca. 48 million). However, for the calculation of the p2p technological similarity score, we only use a subset of all patent applications. First of all, for our present analysis, we only include patent applications which have been granted. We further limit ourselves to patent applications in the period 1980-2017. We further follow De Rassenfosse et al. (2013) and only include priority filings, where we only consider one priority per extended (INPADOC) patent family. Here, we select the earliest priority filing per extended patent family, which by now has been granted and where an English language abstract is available. This leads to a final dataset containing roughly 12 million patent applications.

## 4 Evaluation

In the following, we engage in a first attempt of empirical evaluation of the created technological signature of patents, and the derived p2p similarity measure.

### 4.1 Evaluation strategies

In general NLP research, the evaluation of methods to assess the similarity of text-pairs is a fairly standardized process. In short, such models are usually evaluated on how well they perform on a set of established pre-annotated benchmark "Semantic textual similarity" (STS) datasets (e.g., Bowman et al., 2015, who provide 570k of labeled sentence pairs). While useful for benchmarking models for general language, they tend to not perform well for complex and domain specific language (Chandrasekaran and Mago, 2020) such as technical descriptions to be found in patent text. For patents, currently no STS dataset exists.[5] Guided by recent research on the non-STS based evaluation of text-based vector representation of patents (e.g., Whalen et al., 2020),

---
[5]The only exception is the non-public dataset of patents similarity as labeled by patent experts used by Arts et al. (2018, 2020), where they employ technology experts to label a total of 850 patent-pairs with their similarity. Such a number of pairs is suitable to give initial hint at the promisingness of text-based patent indicators, particularly since they apply a rather outdated methodology. However, for providing a proper benchmark for future improvements of such measure and finetuning of models, a much larger evaluation dataset is needed. While we are actively engaged in developing such a community curated patent STS dataset, these efforts go beyond the scope of the present paper

we apply three complementary strategies: (i.) a "direct" one where we aim at evaluating the created technological signature of patents, and (ii.) a "relational" one where we evaluate the derived patent-to-patent similarity. In section 5, we provide further an (iii.) "indirect" evaluations, where we investigate the plausibility and usefulness obtained results and further derived indicators in a concrete technological setting.

## 4.2 Direct evaluation of the technological signature

The "direct" evaluation assumes that a model can be estimated that links each technological signature to the observable attributes of the corresponding patent. In contrast to a pure classification exercise, we are not *per se* interested in maximising the predictive performance of the technological signatures for classification tasks, but rather their usefulness to measure technological distance.[6] Yet, patent technology classifications contain strong signals regarding the technology embodied in a patent, therefore the created technological signature should enable a better-than-random prediction of the associated IPC classes. This is helpful to assess if the created vector representation contains meaningful information regarding the technology described in the patent.

We in a first step examine whether the produced vectors can perform as inputs for automated IPC symbol classification on sub-class level for the first-mentioned sub-class. This is a multiclass prediction problem with 637 outcome classes in our sample. Using the constructed embeddings to derive indicators requires them being reliable and nuanced representations of the underlying patents. In order to capture interactions and non-linearity between the technological signature and IPC assignments without explicitly modelling them, we deploy an artificial neural network (ANN) with 3 hidden layers, which takes as input the 300-dimensional technological signature vector of a patent and predicts as output the corresponding IPC assignment. Patents can have multiple IPC assignments, making this exercise a multi-class and multi-label prediction problem of predicting all assigned classes. Due to the increased complexity in modelling and evaluation alike, we follow previous research (e.g., Lee and Hsiang, 2020; Li et al., 2018), and for a first evaluation only predict the first mentioned rather than all IPC assignments.[7] We trained the ANN on 9,471,069 observations, and evaluated on 10.0000 out-of-sample observations, which have not been used to fit the prediction model.

---

[6]For embedding-based exercises explicitly aiming at automated patent classification, consider for instance Grawe et al. (2017); Kim et al. (2020); Lee and Hsiang (2020); Li et al. (2018); Risch and Krestel (2019)s.

Table 2: Face validity evaluation of technological signature

| Method | Text data | Data Source | N patents | Target | Level (n class) | Precision | Recall | F1 |
|---|---|---|---|---|---|---|---|---|
| DeepPatent (Li et al., 2018) | title, abstract | USTPO | 2.000.147 | IPC | subclass (637) | 73 | n.a. | n.a. |
|  | title, abstract | EPO, WIPO | 742.097 | IPC | subclass (637) | 45 | 75 | 55 |
| PatBERT (Lee and Hsiang, 2020) | title, abstract | USTPO | 1.950.247 | IPC | subclass (637) | 80 | 64 | 64 |
|  | claims | USTPO | 1.950.247 | CPC | subclass (635) | 84 | 66 | 66 |
| Our approach | title, abstract | EPO | 1.000.000 | IPC | subclass (637) | 54 | 53 | 52 |

*Note: n.a. indicates not reported metrics.*

The classifier achieved a weighted precision of 54%, weighted recall of 53%, and F1 score of 52%, meaning that it was able to detect the right sub-class *out of 637 possible answers* for over half of the patents in the test set. As robustness test, we run a "placebo type" model by shifting all vectors by one observation (relative to the classes). Training and predicting with that setup rendered accuracy and recall values of 0 for nearly all classes. Overall, we conclude that the information created technological signature vectors enables the retrieval of assigned IPC classes reasonably.

Further, more qualitative evaluations can be carried out by providing technological signals and assessing to which extend patents embodying these technologies can be retrieved. To do so, we created a simple application, where a user can enter a free text search string. The application vectorizes the input query using the same language model our technological signature vectors are based on, and returns the patents with the most similar technological signature for visual inspection. We granted a set of domain experts access to this application for testing and evaluation over the period of several weeks. All domain experts reported satisfying performance with respect to the ability of the application to provide them patents embodying the technologies expressed in their search queries. Below, we report the results of an exemplifying search query and the obtained results.

---

[7]Many patent authorities (e.g. USTPO) by law require a patent to be assigned to a main class, which has to be mentioned first. However, for other authorities without such a legal concept, the order of classes is not binding.

**4.3 Relational evaluation of p2p technological similarity**

In the next step we evaluate the comparability of the created technological signatures of patents, and consequently the quality of the calculated p2p technological similarity measure. "Relational" approaches evaluate the similarity measures between two elements (in this case the p2p technological similarity) that can be derived. Conceptually, the validity of the former represents a necessary but not sufficient condition for the latter. In lack of a ground-truth benchmark dataset of annotated p2p similarity, we cannot directly validate how accurate our created measures are. However, we can investigate the correlation between our generated p2p similarity and existing observable measures commonly used to approximate technological similarity. While this exercise as such cannot provide evidence for the advantage of embedding-based p2p similarity measures offer other text based or traditional approaches *per se*, it can serve as a first "sanity-check" of the face validity and plausibility of our results.

Initially, we compare different samples of patent-parts which could intuitively be expected to display on average a higher (lower) similarity, where we rely on the assumption that two patents of the same patent family, developed by the same inventor(s), assigned to the same assignee(s), or that cite each other, are similar to a certain degree. We assume that technological similarity should be more pronounced within technological domains, as approximated by technological classifications such as technological fields, IPC or CPC categories. Following the same argumentation, technological trajectories and specialization of inventors and assignees should lead to a higher similarity of patents filed by the same inventor or assignee as compared to others. Finally, backward citations refer to relevant prior art, therefore we assume a pair where one patent cites the other should on average display a higher technological similarity that pair where this is not the case. We in all cases retrieve all patent-pairs where the respective condition is true (i.e. same IPC class, same inventor/assignee, one patent cites the other), and match it with a random sample of patent-pairs of equal size where this condition is not true. Table 3 reports the results.

Table 3: Face validity evaluation of similarity

| Evaluation       | shared | no shared | t-test |
|------------------|--------|-----------|--------|
| IPC class        | 0.032  | 0.009     | ∗ ∗ ∗  |
| IPC subclass     | 0.049  | 0.010     | ∗ ∗ ∗  |
| IPC group        | 0.071  | 0.010     | ∗ ∗ ∗  |
| IPC subgroup     | 0.108  | 0.011     | ∗ ∗ ∗  |
| inventor         | 0.039  | 0.006     | ∗ ∗ ∗  |
| assignee         | 0.026  | 0.004     | ∗ ∗ ∗  |
| citation         | 0.071  | 0.001     | ∗ ∗ ∗  |
| citation XY      | 0.084  | 0.001     | ∗ ∗ ∗  |
| citation examiner| 0.112  | 0.002     | ∗ ∗ ∗  |

*Note:* Two-sided t-test. $H_1$: True difference in means $\neq 0$.

On average patents within the same IPC class display a significantly higher similarity than patents from different classes. As a result, patents sharing an IPC class display an increased magnitude of similarity by a factor of roughly 3, which increases when repeating the same exercise on subclass (5), group (7) and subgroup (/>9) level.[8] In conclusion, patents sharing at least one IPC subgroup classification are according to our similarity indicator almost ten times more similar than patents which do not. Repeating this procedure on the inventor and applicant level leads to similar results. Likewise, patents filed by the same inventor or assignee are more similar by a factor of roughly 6. All mean differences are significant at the 1% level.

Patent pairs connected by a backward citation show on average a 50 times higher similarity score. However, the average similarity of citing patents is with ca 7% still low. Similar results with slightly higher average similarity and higher correlation are obtained when only limiting ourselves to X and Y tag citations, and citations added by the examiner.[9] While overall reassuring, the outcome is highly skewed, where around 70% of patents citing each other do not display meaningful similarity. Likewise, there are many patent pairs with high similarity scores that do not cite each other, supporting previous findings regarding the bias associated with citation data (e.g. Alcacer and Gittelman, 2006; Bacchiocchi and Montobbio, 2010; Lampe, 2012; Picard and de la Potterie, 2013), and the conclusion that citations may provide only a limited indicator technological similarity.

---

[8] Similar results are obtained when using the CPC classification scheme instead. Sharing multiple classes further increases our similarity score.

[9] Likewise, the Pearson correlation coefficient between citation and similarity of a patent pair is with 0.05 low but statistically significant at the 1% level.

Finally, we compared the performance of our similarity calculation to Arts et al. (2020). We construct and perform approximate nearest neighbor search with the sample of over 6 million USPTO patents for which the authors published their similarity measures in a data repository. We then compare the overlaps in IPC assignments for the most similar 10 patents identifying how many of those identified as similar are sharing IPC assignments on class, subclass and group level. Table 4 below presents the share of patents with at least one overlap on different levels. Our results are on all levels a few percent points under Arts et al. (2020) indicating that both approaches are capturing very similar features. Given the nature of the embeddings we propose that is tuned to capture synonyms, we can speculate – following McNamee (2013) – that more patents are identified as similar where the technology with a different IPC assignment is similar in terms of its application but does not share a similar terminology.

Table 4: Comparison similarity calculation vs Arts et al. (2020)

|                | class level | subclass level | group level |
|----------------|-------------|----------------|-------------|
| Arts et al. 2020 | 0.3855106 | 0.2806797    | 0.1307425   |
| Our approach   | 0.3639752   | 0.2442472      | 0.1007457   |

To exemplify the results of our semantic search queries, we picked random patents from the e-mobility field and searched for the most similar patents in our patent database. We found that similarity is not always reflected by an overlap of IPC classes but often better by similarity of technical descriptions. Moreover, sometimes a high degree of similar can also be found even though different vocabulary is used. Table 5 shows examples of technical descriptions from picked patent abstracts and extractions of descriptions from most similar patents as found by the search query. As illustrated by the results, we obtain matches that resemble the search string in terms of meaning without necessarily the need for exact keyword matches. This indicates capability of the presented method to identify technologically related patents to text queries of arbitrary length, and thereby the usefulness form applications such as semantic search and patent retrieval.

Table 5: Comparison similarity calculation

| Original | Most similar |
|---|---|
| "The invention relates to a lateral guidance control structure with one or more control variables for generating a steering input of a power steering system of a motor vehicle. A device for controlling lateral guidance of a vehicle is described." | "The invention relates to a device for operating a servo steering system which has at least one electric motor for generating a supporting motor steering torque for pivoting at least one steerable wheel of a vehicle." |
| "The invention relates to a method and a device for detecting a lane change in the context of a vehicle speed control system, such as ACC (Automatic Cruise Control) or a system for distance or collision warning in vehicles, in which a lane change mode is activated as a function of a lane change probability." | "Purpose an automatic driving system for detouring danger area in the automatic driving of a vehicle is provided to control progressive direction of the automatic traveling vehicle by using the reference direction information, 'constitution an automatic driving system for detouring danger area in the automatic driving of a vehicle comprises a road lane recognition part a car speed detecting part a train compartment distance detection part a gps signal receiving part a wireless communication unit a memory unit a driving controller and an automatic driving control unit" "The invention discloses an intelligent lane changing assisting system for an intelligent vehicle and a control method thereof which belongs to the field of automobile active safety." |
| "The invention relates to a method for environment detection of a vehicle in current driving situations, in which objects are detected and tracked from the environment and the detection and tracking of these objects is performed with an adjustment of the environment sensors executed as a function of the vehicle state." | "Monitoring device for a motor vehicle has a sensor device for detecting obstacles in front of or behind the vehicle and an evaluation unit for checking for the presence of an obstacle within a predefined distance from the vehicle within a monitoring surface." "The device has a vehicle camera provided as an environment sensor and an evaluation unit for determination and output of data based on lights of a vehicle to be automatically controlled." |
| "The invention relates to a control system for a vehicle with actuators." | "Problem to be solved to provide a control system for controlling a main electronically controlled vehicle system and further controlling at least one additional auxiliary vehicle system." "A vehicular electronic control apparatus includes a vehicle control means and a unit control means." |
| "The invention relates to a driving style evaluation device. A driving behavior representation parameter estimation unit provides an estimated value of a driving behavior representation parameter representing the driving behavior of the driver of a vehicle." | "The invention relates to a method for operating at least one motor vehicle said method involving the steps of providing s a data record characterizing a vehicle environment a driving behavior and the movement of the motor vehicle ascertaining s at least." "The present invention relates to an arrangement and a method for estimating the speed of a vehicle." "The method includes but is not limited to the steps of evaluating a driver s driving style." |

# 5 Case Study and Research Applications: Electromobility Patents

While the evaluation of basic characteristics is helpful to ensure certain properties such as plausability, an analysis on a limited and somewhat heterogeneous set of technologies enables us to derive context-dependent insights based on domain expertise. At the case case of electric vehicle (EV) technologies we provide a more contextual illustration of the properties of the proposed text-based technological signatures and p2p similarity, and showcase promising research applications.

In particular, we demonstrate the use of p2p similarity indicators and illustrate the obtained results for two popular research applications: i.) to create patent quality indicators, and ii.) to map cross-country knowledge flows. Acknowledging the vast body of research on both topics, we do not claim these stylized applications to advance both lines of research as such. Rather, they aim at providing examples of where and how p2p similarity measures can be used, and provide intuition about their outcomes within a well defined technology case.

## 5.1 Context and data

EV technology is currently about to develop from the niche into the mass market. Thereby, it fosters a shift in the technological regime, leaving the internal combustion engine (ICE) technology behind. Apart from being more environmentally friendly, electric vehicles have a number of further advantages: "Electric motors are low-maintenance, versatile and exceptionally quiet" (p. 4 Deffke, 2013).

We identify EV technologies using IPC codes on the subclass level,[10] focussing on electric propulsion, a key technology of battery electric vehicles (BEVs). To identify EV related patents, we follow Pilkington and Dyerson (2006) and select patents to be found in the IPC class B60L 11/00 and its subclasses, as they can be determined as a "likely home for EV patents" (Pilkington and Dyerson, 2006, p. 85).[11] We analyse

---

[10] Whereas group and subgroup labels allow even more nuanced identification, they are also less stable over time due to more frequent revision, addition, and reclassification (WIPO, 2017).

[11] A list of all used IPC-classes and their description is given in Table 6 in the appendix. Figure 5

priority patents granted in the the period 1980-2012 assigned to the associated IPC classes, which results in 22,285 patents.

## 5.2 Research application 1: Patent quality indicators

It has long been recognized that the technological as well as economic significance of patents varies broadly (Basberg, 1987), and as a result a large body of literature has explored the rich information contained in patent data to construct patent quality measures.[12] Such measures are traditionally constructed based on (i.) the number or composition of assigned IPC classes (e.g. Lerner, 1994), (ii.) the number and pattern of backward citations (e.g. Harhoff et al., 2003a; Lanjouw and Schankerman, 2001; Schoenmakers and Duysters, 2010) or (iii.) forward citations (e.g. Ahuja and Lampert, 2001; Hall et al., 2005; Harhoff et al., 2003b), and (iii.) IPC class composition of citing or cited patents (e.g. Shane, 2001; Trajtenberg et al., 1997,?; Uzzi et al., 2013). Among those, many suggested measures of different aspects of patent quality are explicitly or implicitly based on the their similarity to other patents, particularly the ones published at earlier or later points in time. Patents with a high similarity to earlier patents are assumed to build on existing knowledge, technologies, and applications, whereas low similarity to earlier work indicates *novelty* (Arts and Veugelers, 2015; Uzzi et al., 2013). Patents with a high similarity to patents published after the focal one indicate the *promisingness* of the embodied technology, since it will frequently be applied in the future. Recently, first attempts to create text-based patent quality measures leveraging p2p similarity have been made, primarily leveraging simple (Arts et al., 2018) or TFIDF weighted (Arts et al., 2020; Kelly et al., 2021) co-occurrence. Indeed, text-and similarity-based measures of patent quality appear to correlate well with a large array of ex-post quality measures, such as patent value (Kelly et al., 2021) and the association with prestigious technology awards (Arts et al., 2020).

However, traditional as well as text-based ex-ante quality measures are found o vary

---

and Figure 6 provide an adittional visualization of rechnological relationships between these IPC classes.

[12] For a somewhat recent and exhaustive review on patent quality measures, consider Squicciarini et al. (2013), and for a more critical reflection on them Higham et al. (2021).

substantially with respect to different post-grant outcomes associated with patent quality, and display significant variation within the same measure across technologies within outcomes (Higham et al., 2021). In short, patent quality is an ongoing field of study, and while text- similarity- based indicators appear promising, there exists no consensus on how to construct them and based on which particular datasource, particularly since they tend to be sensitive to the outcome of interest as well as variations between technologies. Without claiming to provide a superior approach, we suggest that our embedding-based p2p similarity measure can be used to complement and augment existing approaches, since embeddings are less sensitive to domain-specific technical jargon of particular technology fields. We in the following provide a simple approach to leverage embedding-based p2p similarity to construct two popular measures of patent quality, technological *novelty* (lack of similarity to earlier applications) and impact (the similarity to later applications), which can be used as point-of-departure for future indicator development.

To do so, we first sum all the similarity relationships a patent displays to the universe of other patents, resulting in indicator $sim_i$. For every patent $i$, $J_i[1:m]$ will contain patents $j$ with earlier as well as later application dates. This difference is measured in years with the parameter $\Delta t_{j,i} = t_j - t_i$. With that information, we can construct a temporal similarity index on patent level, which captures its similarity to other patent applications filed earlier ($sim_i^{past}$) or later ($sim_i^{future}$).

$sim_i = \sum_{j=1}^{m} \frac{s_{i,j}}{m}$

$sim_i^{past} = \sum_{j=1}^{m} \frac{(-\tau > \Delta t_{j,i} \geq -\lambda) \ s_{i,j}}{m}$

$sim_i^{future} = \sum_{j=1}^{m} \frac{(\tau < \Delta t_{j,i} \leq \lambda) \ s_{i,j}}{m}$

The resulting indicators represent $i$'s share of similar patents with application date in the past ($sim_i^{past}$) or future ($sim_i^{future}$), weighted by their similarity $s_{i,j}$. ($-\tau > \Delta t_{j,i} \geq -\lambda$) is a logical condition, leading to the inclusion a multiplier for $s_{i,j}$ of one or zero depending if the condition is fulfilled or not. To offset the delay between patent application and its official publication of up to 12 months (Squicciarini et al., 2013), we introduce a parameter $\tau$ that represents minimum $\Delta t_{j,i}$ for $j$ to be considered in the temporal similarity indicators (we here set $\tau = 1$). In addition, the parameter $\lambda$ restricts the maximum $\Delta t_{j,i}$ for patent $j$ to be included in the calculation of the temporal similarity indicators. To make it consistent and comparable with the traditionally used 5-year forward citation count as patent quality indicator (eg. Harhoff et al., 2003a; Squicciarini et al., 2013), we set $\lambda = 5$.

At the case of EV technologies, we illustrate and discuss obtained results.[13] For a first overview over the sector and technology, Figure 2 displays the development of the number of EV patent applications as well as their average $sim_n^{future}$ over time.

Figure 2: Overal number and similarity of EV patents

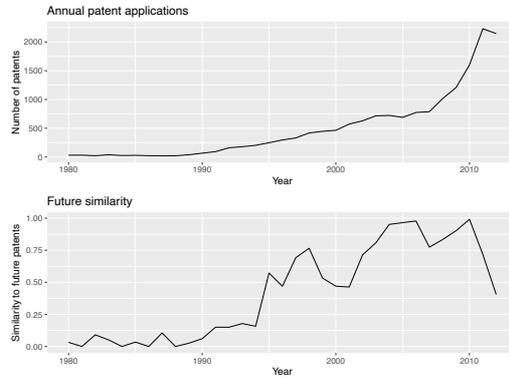

While we see marginal activity in patent applications already in the 1980s, we only see a steady growth beginning in the 1990s, with a sharp increase in the mid-2000s. $sim_n^{future}$, however, follows a different trajectory. Until the mid-1990s, almost no patent showed similarity to future patents, indicating the generally low patenting activity but also the non-cumulative and fragmented nature of technology development in this period. However, in the mid 1990s, we witness a sudden peak of $sim_n^{future}$, followed by further peaks in the mid-2000s and early 2010s, which hints at an by now visible technology life-cycle. Here, the first main peak around the year 1997 coincides with the development of the Toyota Prius, becoming the first mass-produced hybrid-electric vehicle and forerunner in the field of (hybrid) electric vehicle technology. The following peaks fall into the time of growing patenting activity in the energy storage field in general but with a steady rising focus especially on lithium-ion technologies in 2005 and the following years (Dinger et al., 2010). This technology played an important role in electric vehicle development within the next decade, which explains the high future similarity in this time. The figure also illustrates the forward-looking nature of $sim_n^{future}$, where new technological trends and developments are traceable before the corresponding technology starts enjoying its popularity. Consequently, we suggest that on different levels of aggregation, $sim_n^{future}$ can be interpreted as an indicator of the "impact" of certain technology.

---

[13] Further examples of similar applications can be found in Hain et al. (2020), where we utilize p2p to measure technological catching-up efforts on country level.

In the following, we provide an overview of EV patenting and our similarity-based indicators on country-level.[14] 3 illustrates the technological development of the five countries accounting for the highest number of EV patent applications, Japan, South Korea, the United States, Germany, and France.

Figure 3: Novelty & Impact on country level

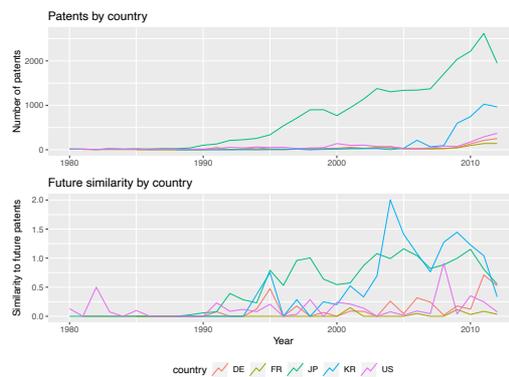

Based on the displayed curves Japan can clearly be identified as the leading country in the field of core electric vehicle patents showing a sharp increase in output since the 1990s as the general forerunner in EV technologies. This is in accordance with the development of Japans vehicle industry, which was the first to introduce vehicles with alternative powertrains and was also strongly supported by governmental programmes at an early stage (Åhman, 2006). This first position remained unchallenged for the whole period considered. However, in the mid-2000s, Korean EV-research started to take off and increased its patent output rapidly thereafter. This uptake is clearly in line with and the founding of the Pangyo Techno Valley (PTV) in 2004, a large research cluster accumulating eight of the top 10 Korean tech companies and more than 1,300 IT-companies as well as the introduction of the Korean "Innocity" policy in 2007 to establish new innovation cities (Lee et al., 2017). The United States, Germany, and France, in the meantime, showed only negligible activity and just around 2010 became somewhat significant. This possibly results from a comparatively late introduction of EV-innovation policies for the US in 2009 (Gu and Shao, 2014) and the PPP Green Car Initiative of the European Commission starting in 2008. Overall, patenting in EV technologies appears as rather concentrated, where Japan accounts for 41% of all patents filed, and the leading five countries are together the creators of 89% of all patents. In terms of the development of the $sim_n^{future}$ indicator, we see a slightly different picture compared to the total patent count. The development of Japan's $sim_n^{future}$ roughly follows its amount of patent applications and shows a somewhat stable trend of high future similarity. The huge impact of the Japanese patent count can also be seen in the resemblance of the Japanese course to the overall similarity in Figure 3. However, we also spot several peaks of countries with at that point in time minimal patenting activity, but promising technologies developed. Particularly noticeable is the peak of South Korea in the mid-2000s, where the average $sim_n^{future}$ of Korean patents overtakes Japans lead.[15] Overall, the high average future similarity of Korean patents in the following years of rising patent count suggests a highly innovative and future driven patenting behavior.

---

[14]PATSTAT data is known to incompletely capture inventor addresses correct and complete (ca. 30% of patents cannot be clearly assigned to any geographical location), a problem which is amplified particularly in Asian countries. Therefore, for this research, we leverage recent efforts by De Rassenfosse et al. (2019) to provide more comprehensive geo-information for PATSTAT data, covering more than 90% of global patenting activity. Since most patents have multiple inventors listed, we assign every geolocation a fractionalized number representing the share of inventors of a particular patent in a particular location. We choose the inventor information instead of the more commonly used applicant information to assign patents to countries in order to capture the location of inventive activity rather than the location of intellectual property right ownership (Squicciarini et al., 2013).

---

[15] However, this peak is mainly caused by the big differences in patent count among countries at this time and a graph that is based on country averages, as the most promising patent from Japan still ranks three times higher in future similarity than the best Korean one.

## 5.3 Research application 2: Mapping knowledge flows

The introduced temporal p2p similarity indicators naturally lends itself to a direct network analysis on different levels of aggregation. As an example, we in the following create a directed network between the top-patenting countries based on aggregated $sim_{i,j}^{future}$. Since the similarity of patent applications in country $i$ with patent applications in country $j$ at a later point in time can be interpreted as a knowledge spillover, the resulting network illustrates technology related knowledge flows between countries.

Figure 4: Knowledge flows between countries

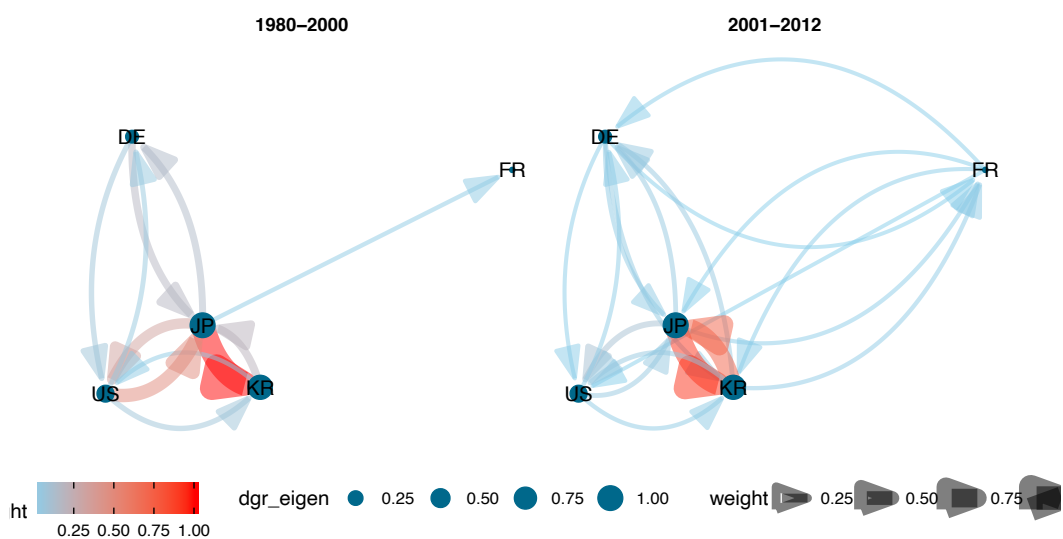

Note: Directed network based on $sim_{i,j}^{future}$ on country level. Node size reflects the in-degree eigenvector centrality Edge color and weight represents edge weight.

We see that during the formative period of EV technologies until 2000, strong knowledge flows particularly from Japan to South Korea, but also bidirectional ones between the US and Japan can be observed. The network underlines that Japan can be seen as the central player in this time, building the knowledge base for future developments of the other top 4 patenting countries. However, it also becomes apparent, that some Japanese developments in turn base on US-American developments in the early 80's, mainly a patent introducing LiCoO2 as a new cathode material for lithium batteries ([Godshall et al., 1982](#)). The strongest knowledge flow for the first period is observable from Japan to Korea, which goes along with the dissemination of knowledge in other technological fields. Former studies showed, that for several high technological areas like flat panel displays (FPD) ([Hu, 2008](#); [Jang et al., 2009](#)) as well as the mobile telephones ([Lee and Jin, 2012](#)) the knowledge source / patent citation often follows the order of industry entry leading to Japan following the US and Korea following Japan ([Han and Niosi, 2018](#)). Further reasons for the strong connection might also be seen in the high resource-based dependence from Japan, with for example LG Chem, the largest Korean EV-battery producer, being heavily reliant on Japanese materials.

Besides an overall higher connectivity of the knowledge flow network post 2000, also its characteristics changed. Beside this apparent increased interconnectedness between the countries, we now see strong bidirectional connections between Japan and South Korea, indicating mutual reinforcing knowledge flows. Conversely, knowledge flows between the US and Japan now mainly originate from Japan.

## 6 Conclusion

In this paper, we propose an efficient and scalable approach to create vector representations of a patent's technological signature based on textual information to be found in their abstract by utilizing embedding techniques from natural language processing. We leverage these technological signatures to derive p2p technological similarity measures. We suggest and demonstrate the use of approximate nearest neighbor matching to create similarity measures for a large datasets, allowing us to represent the whole universe of patents as a similarity network and thereby opening the possibility for a large range of applications and analyses. We evaluate the properties of our embedding-based p2p similarity indicator in various ways, illustrate obtained results and suggest potential research applications at the case of electromobility technologies.

While the results so far demonstrate the usefulness of a semantic indicator of p2p technological similarity, and give a first glance at possible applications, the full potential remains somewhat unexplored. In the following, we indicate what needs to be done in order to improve the accuracy of technological signature and the derived p2p similarity measures, to validate its outcome, and apply it to a range of suitable problems.

First, the present approach is utilising patent abstracts as a data source, relying on them being rather standardised summaries of the technologies described by the patents. This allows to apply existing NLP approaches without violating underlying assumptions of the computational models used, about comparable length of texts across a corpus or non-ambiguity of text fragments. Future research should – following the current developments in language processing technologies – however explore inputs beyond abstracts utilising patent claims and eventually full texts. To do so, an increased understanding which sources of textual data contain extractable information on the technology, application, novelty, or legal protection, and how different datasources can be combined to provide more holistic representations of a patent's technology.

With respect to the validity, information content, and use of the patent's technological signature, several avenues for future work exist. Based on our evaluation, we are confident that the vectors do represent the underlying patents' technological features, since it enables the prediction of the patent's technology class. Generally, the validation and verification of the proposed measure of technological similarity between patents is limited to the reproduction of stylized facts and the comparison to existing measures. While first attempts to utilize domain expert knowledge to validate and optimize technological similarity metrics have been made (Arts et al., 2018), the creation of a large-scale expert annotated dataset could create an objective benchmark, allowing technology forecasting researchers. Guidance here can be drawn from large pre-annotated "Semantic textual similarity" (STS) datasets frequently used in natural language processing research.

# Appendix

Table 6: List of used IPC-classes

| IPC class | Level | Description |
|---|---|---|
| B60L 11/00 | Subgroup | Electric propulsion with power supplied within the vehicle |
| B60L 11/02 | Subgroup | Using engine-driven generators |
| B60L 11/04 | Subgroup | Using dc generators and motors |
| B60L 11/06 | Subgroup | Using ac generators and dc motors |
| B60L 11/08 | Subgroup | Using ac generators and motors |
| B60L 11/10 | Subgroup | Using dc generators and ac motors |
| B60L 11/12 | Subgroup | With additional electric power supply, e.g. accumulator |
| B60L 11/14 | Subgroup | With provision for direct mechanical propulsion |
| B60L 11/16 | Subgroup | Using power stored mechanically, e.g. in flywheel |
| B60L 11/18 | Subgroup | Using power supplied from primary cells, secondary cells, or fuel cells |

Figure 5: UMAP projection of patent vectors

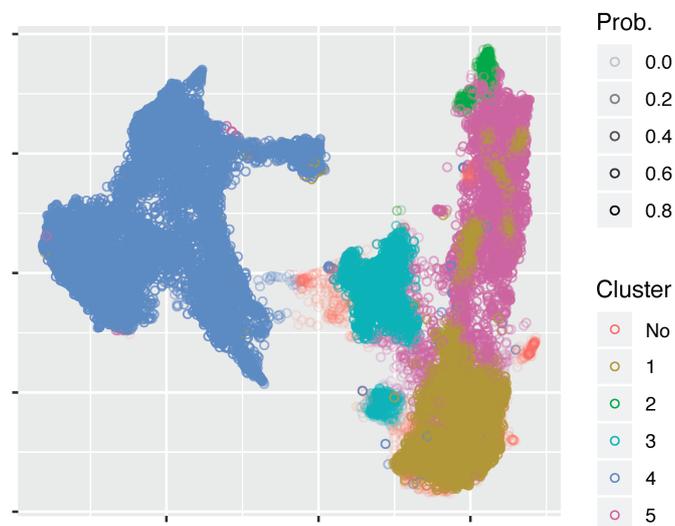

Note: UMAP dimensionality reduction (McInnes et al., 2018) of EV patent signatures in 2-dimensional space. Colors indicate the outcome of a density-based clustering (HDBSCAN).

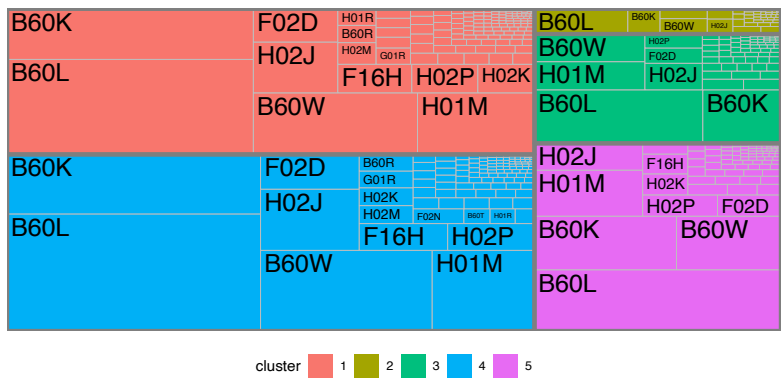

Figure 6: IPC class composition of technology clusters

Note: Illustration of IPC composition of identified clusters in EV patents.

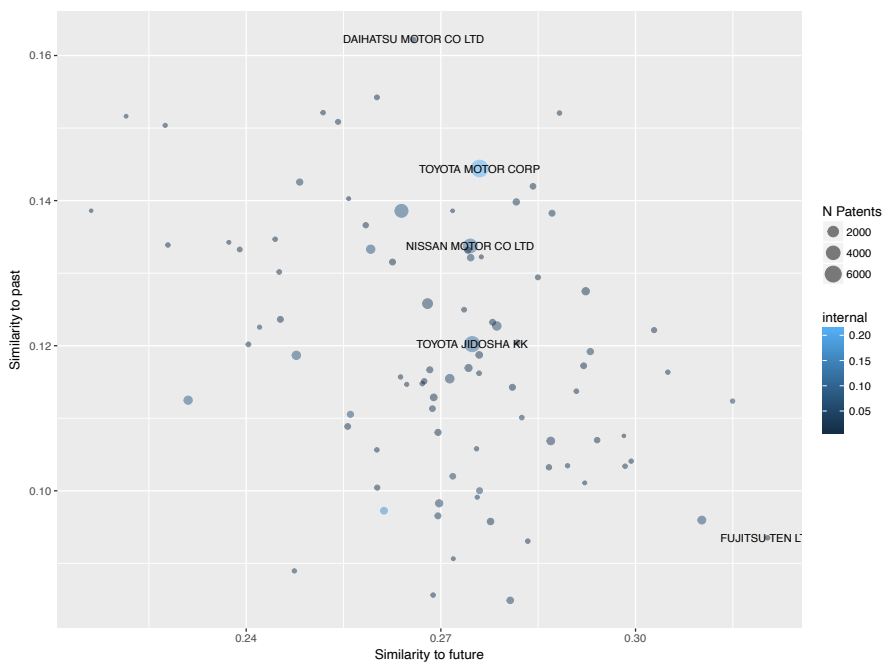

Figure 7: Novelty & Impact on firm level

45